\documentclass[aps, prx,twocolumn]{revtex4-2}  

\usepackage[utf8]{inputenc}
\usepackage[colorlinks=true, linkcolor=blue, urlcolor=blue, citecolor=blue]{hyperref} 
\usepackage{url}

\usepackage{tikz} 
\usetikzlibrary{quantikz2}

\usepackage{amsmath}
\usepackage{amssymb}

\usepackage{graphicx}

\usepackage{orcidlink}

\begin{document}

\title{Schr\"odinger-Navier-Stokes Equation for the Quantum Simulation of Navier-Stokes Flows}

\author{Luca Cappelli \orcidlink{0009-0009-1169-8380}}
\email{luca.cappelli@iit.it}
\affiliation{Fondazione Istituto Italiano di Tecnologia, Center for Life Nano-Neuroscience at la Sapienza, Viale Regina Elena 291, 00161 Roma, Italy}

\author{Sauro Succi \orcidlink{0000-0002-3070-3079}}
\affiliation{Fondazione Istituto Italiano di Tecnologia, Center for Life Nano-Neuroscience at la Sapienza, Viale Regina Elena 291, 00161 Roma, Italy}

\author{Monica L\u{a}c\u{a}tu\c{s} \orcidlink{0009-0006-0827-1800}}
\affiliation{Delft Institute of Applied Mathematics, Delft University of Technology, Mekelweg 4, Delft, 2628 CD, Zuid-Holland, The Netherlands}

\author{Alessandro Andrea Zecchi \orcidlink{0009-0000-4091-768X}}
\affiliation{MOX – Department of Mathematics, Politecnico di Milano, Piazza L. da Vinci, 32, 20133 Milano, Italy}

\author{Alessandro Roggero \orcidlink{0000-0002-8334-1120}}
\affiliation{Dipartimento di Fisica, University of Trento, via Sommarive 14, I38123, Povo, Trento, Italy}
\affiliation{INFN-TIFPA Trento Institute of Fundamental Physics and Applications, Trento, Italy}

\begin{abstract}
The search for quantum-like wave formulations of the Navier-Stokes 
(Schr\"odinger-Navier-Stokes, SNS for short) equations describing
classical dissipative fluids has met with increasing attention in 
the recent years, due to the large portfolio of potential applications 
in science and engineering 
\cite{Tennie2025, SREENI, MengZhongXu2024_SimulatingUnsteadyFlows}. 
A SNS formulation of classical fluids was first presented in a 
largely un-noticed paper by Dietrich and Vautherin back in 1985~\cite{dietrich_vautherin}. 
In this paper, we revisit this specific SNS approach and assess its viability
for quantum implementations based on Carleman embedding/linearization 
techniques. 
Specifically, we
i) Clarify in full mathematical detail why the SNS dissipator presents a steep
challenge for quantum computers and propose a way out strategy 
based on the Hamilton-Jacobi (HJ) formulation of fluid dynamics; 
ii) Develop a corresponding quantum algorithm using a new technique 
based on a tensor-network representation of Carleman embedding of 
the HJ equations (CHJ) which permits substantial memory savings;
iii) Emulate the CHJ quantum algorithm on a classical computer
and analyse its convergence and accuracy for the specific case
of Kolmogorov-like flows at moderate Reynolds numbers.

To the best of our knowledge, this is the first quantum algorithm based 
on a quantum-like wave formulation of the genuine Navier-Stokes 
equations, including pressure, dissipation and vorticity.
\end{abstract}

\maketitle

\section{Introduction}

The recent years have witnessed a mounting interest towards the quantum
simulation of classical physics problems, notably transport 
phenomena~\cite{SANAVIO_ADR, diMolfetta2024, Schumacher24_ADR, BharadwajSreenivasan2025} and
the physics of fluids~\cite{Tennie2025, diMolfetta2022, wiebe25, psiq_26, LacatusMoller2026, Itani2022}. 
Besides the usual challenges of decoherence and loss
of entanglement, the quantum simulation of classical fluids presents two major additional
hurdles, {\it Nonlinearity} and {\it Dissipation} ~\cite{EPL2023}. Various strategies have been proposed to
handle these challenges, among others the so-called hydrodynamic Schr\"odinger formulation,
a generalization of the well-known Madelung formulation of quantum mechanics, whereby 
the hydrodynamic equations are cast within a quantum-like wave formalism~\cite{SALAS, MengZhongXu2024_SimulatingUnsteadyFlows, MengYang2024_QSR_NS}. 
The rationale behind this approach is that the formal analogy with quantum 
mechanics should facilitate the formulation and implementation of the 
corresponding quantum algorithms.
The transition from the genuinely quantum-mechanical Madelung fluid to classical
Navier-Stokes (NS) fluids involves three major steps: 1) replace the non-local quantum 
potential with classical pressure, 2) account for  dissipation and 3) vorticity.
In \cite{SALAS}, the authors indicate how to cope with 1) and 2) but leave 3) hanging. 
In \cite{MengZhongXu2024_SimulatingUnsteadyFlows}, a spinorial wavefunction, coupled 
to a magnetic field, is introduced to account for vorticity. 
This theoretical formulation is elegant but mathematically intense and it suffers from
an inconsistent representation of dissipative effects, which are surrogated 
by a sign-changing artificial force. Such limitation has been  
removed in a subsequent paper \cite{MengYang2024_QSR_NS},
but the treatment of the correct Navier-Stokes dissipator is handled by a purely classical
algorithm.  Hence, the problem of formulating a viable quantum SNS algorithm remains open.
Quoting verbatim from the original source " ... devising an effective quantum algorithm 
for evolving this quantum system remains a valuable open problem".

In the present paper we bridge this gap in three main respects. 

First, we note that a hydrodynamic wave equation compliant with the three requirements
1-3) above, was already available since the early 80's \cite{dietrich_vautherin}. 
Such wave equation looks mathematically simpler and physically more transparent 
than the spinorial formulation, as it relies upon a {\it scalar} wavefunction 
coupled to a magnetic field. 
Second, by making use of tensor-network formalism~\cite{Verstraete2008_MPSReview, Orus2014_TensorNetworks}, we develop a systematic 
algorithm based on Carleman linearization of the above hydrodynamic 
wave equation in Hamilton-Jacobi form, CHJ for short.
Third, we emulate the quantum algorithm on a classical computer and show that it
performs better than Carleman linearization of the Navier-Stokes 
equations (CNS)~\cite{Sanavio_2024}, and comparably  with Carleman linearization of the lattice Boltzmann 
formulation (CLB) \cite{sanavio_CLB}.     
  
\section{The Navier-Stokes Schr\"odinger equation}\label{sec:the starting equation}
The Madelung formulation of quantum mechanics  \cite{Madelung1926} 
casts the  Schr\"odinger equation (SE) into a  hydrodynamic framework 
through a nonlinear change of variables, expressing the complex 
wavefunction $\psi(\mathbf{r},t)$ in terms of mass density $\rho(\mathbf{r},t)$ and 
a scalar velocity potential $\chi(\mathbf{r}, t)$
\begin{equation}\label{eq:madelung}
    \psi(\mathbf{r},t) = \sqrt{\frac{\rho(\mathbf{r},t)}{m}} \exp \left\{ \frac{im}{\hbar} \chi(\mathbf{r},t)\right\} \, ,
\end{equation}
where the associated flow field is defined 
by $\mathbf{v}(\mathbf{r}, t) = \nabla\chi(\mathbf{r}, t)$.

The Madelung flow carries around the quantum probability density $\rho(\mathbf{r},t)$
according to the laws of quantum mechanics.
As a result, it differs drastically from ordinary classical fluids: it is 
inviscid, irrotational and subject to a non-local quantum potential 
(quantum fluctuations) rather than the ordinary pressure (classical fluctuations). 

The fluid description to the Schr\"odinger equation invites  
a natural question: is it possible to define a quantum-like 
wave equation exactly equivalent to the Navier-Stokes equations (NSE)?
This construction is commonly referred to as the inverse Madelung transformation.
The inverse Madelung transform bears a fundamental interest in terms
of shedding new light into the long-standing problem of the emergence
of classical behaviour from the underlying quantum world~\cite{inverse_madelung}.

However, in the recent years, the search for such a formulation has attracted considerable 
attention as a potential route to the quantum simulation of classical fluids~\cite{SALAS, MengZhongXu2024_SimulatingUnsteadyFlows, diMolfetta2022}.
To this end, several models have been proposed, the most recent one 
based on the Schr\"odinger-Pauli equation~\cite{MengYang2024_QSR_NS}. 
Nevertheless, back in 1985 Dietrich and Vautherin~\cite{dietrich_vautherin} proposed 
a simpler solution based on the standard Schr\"odinger equation, in which the 
rotational component of the associated fluid emerges from the coupling 
to a rotational external field $\mathbf{A}(\mathbf{r},t)=(A_x(\mathbf{r},t),A_y(\mathbf{r},t),A_z(\mathbf{r},t))^T$, whose curl
$$
\mathbf{\omega} = \nabla \times \mathbf{A}
$$ 
represents the fluid vorticity.
This mechanism closely resembles the dynamics of a  
charged particle spinning around the magnetic field. 
The hydrodynamic Schr\"odinger equation reads as follows:
\begin{align}\label{eq:schrodinger}
    i\hbar \frac{\partial \psi}{\partial t}
    = &
    \Big\{
    \frac{1}{2m}\left(\mathbf{p} + m\mathbf{A}\right)^2
    + U + 
    \\ & \nonumber
    + (\mu + \nu)
    \left[
    m\,\nabla \cdot \mathbf{A}
    - \frac{\hbar}{2i}\,\nabla^{2}
    \ln\!\left(\frac{\psi^{*}}{\psi}\right)
    \right]
    \Big\}
    \psi \,,
\end{align}
where $\nu$ and $\mu$ are constant viscosity coefficients. 
This corresponds to assuming that the viscosities depend linearly on the density,
\begin{equation}\label{eq:hp_on_viscosity}
    \xi(\mathbf{r}, t) = \mu \, \rho(\mathbf{r}, t), \qquad 
    \eta(\mathbf{r}, t) = \nu \, \rho(\mathbf{r}, t),
\end{equation}
so that the coefficients remain uniform throughout the flow. 
The potential term $U(\mathbf{r},t)$ is then defined as
\begin{equation}\label{eq:potential_u}
\begin{split}
    U(\mathbf{x},t) &= m h(\mathbf{r}, t) + Q(\rho(\mathbf{r}, t)) \\
    &= m h(\mathbf{r}, t) 
    + 
    \frac{\hbar^2}{2m}\frac{\nabla^2 |\psi|}{|\psi|} \,,
\end{split}
\end{equation}
were $h(\rho(\mathbf{r},t))$ represents the enthalpy per unit mass under the barotropic hypotheses: $P=P(\rho(\mathbf{r}, t))$. Under this condition the enthalpy  can be found by solving the equation
\begin{equation}\label{eq:enthalpy_h}
    \nabla h = \frac{1}{\rho}\nabla P \,.
\end{equation}
The barotropic closure allows the pressure-gradient term in the NS equation to be rewritten in a scalar form directly comparable to the hydrodynamic system obtained from the Madelung transformation.
 \\
An additional quantum pressure term $Q(\rho(\mathbf{r},t))$ is introduced 
to cancel the quantum Bohm potential, an intrinsic feature of the SE, emerging 
from the Madelung formulation, with no counterpart in the NSE.
Finally, provided that the external field \(\mathbf{A}(\mathbf{r},t)\) evolves according to
\begin{equation}\label{eq:evolution_of_A}
    \frac{\partial \mathbf{A}}{\partial t} = \mathbf{v} \times \left(\nabla \times \mathbf{A}\right) - \nu\,\nabla \times \left(\nabla \times \mathbf{A}\right) \,,
\end{equation}
the SNS in Eq.~\eqref{eq:schrodinger} describes a viscous fluid with 
non-vanishing vorticity, where the velocity field
\begin{equation}\label{eq:definition_v}
    \mathbf{v}(\mathbf{r}, t)= \nabla \chi(\mathbf{r}, t) + \mathbf{A}(\mathbf{r}, t)
\end{equation}
is decomposed into an irrotational component and a rotational one, the latter 
being encoded in the vector field \(\mathbf{A}(\mathbf{r},t)\).
For the sake of self-containedness, in the following we recall the main ideas
behind the Dietrich-Vautherin derivation.

\subsection{Wave representation of Navier-Stokes fluids}

Consider the complex wavefunction expressed in the Madelung 
formalism~\eqref{eq:madelung}, where $\rho(\mathbf{r},t), \chi(\mathbf{r},t)$ are real-valued field. 
Substituting this expression for the wavefunction in Eq.~\eqref{eq:schrodinger}, and 
splitting the real and imaginary part of the resulting equation, we obtain 
the continuity equation for $\rho(\mathbf{r},t)$ and a 
Hamilton-Jacobi-like equation (HJ) for the phase $\chi(\mathbf{r},t)$. 

From the real part we obtain:
\begin{equation}\label{eq:continuity_chi_A}
    \partial_t \rho + \nabla \chi \cdot \nabla \rho + \mathbf{A} \cdot \nabla \rho + \rho (\Delta \chi + \nabla \cdot \mathbf{A}) = 0 \,;
\end{equation}
while the imaginary part becomes
    \begin{align}\label{eq:HJ_chi_A}
        \partial_t \chi 
         + \frac{|\nabla \chi|^2}{2} + \nabla \chi \cdot \mathbf{A} + \frac{|\mathbf{A}|^2}{2} + h(\rho) &
        \\\nonumber
        - (\mu + \nu)(\Delta \chi + \nabla \cdot \mathbf{A}) = 0 &   \,.
    \end{align}

Using the expression for the velocity field~\eqref{eq:definition_v}, 
Eq.~\eqref{eq:continuity_chi_A} reads as the continuity equation, while 
Eq.~\eqref{eq:HJ_chi_A} as an Hamilton-Jacobi-like equation for the phase $\chi(\boldsymbol{r}, t)$.
\begin{align}\label{eq:continuity}
    & \partial_t \rho + \nabla \cdot (\rho \, \mathbf{v}) = 0 \, ;
    \\ \label{eq:Hj}
    & \partial_t \chi + \frac{\boldsymbol{v}^2}{2} + h(\rho) - (\mu + \nu) \, \nabla \cdot \boldsymbol{v} = 0    
\end{align}
Finally, combining Eqs.~\eqref{eq:evolution_of_A},~\eqref{eq:definition_v},~\eqref{eq:continuity} and~\eqref{eq:HJ_chi_A} one obtains the Navier-Stokes equations (NSE) in Lagrangian form
\begin{equation}
    \label{eq:NSE}
    \frac{\partial\mathbf{v}}{\partial t} + (\mathbf{v\cdot\nabla})\mathbf{v}
    =
    -\nabla h + \mu \nabla(\nabla \cdot \mathbf{v}) + \nu  \nabla^2\mathbf{v} \,.
\end{equation}

For weakly compressible flows (see below), this can be turned into a cubic one
by expanding around a reference density $\rho_0=1$,   
$\rho^{-1} \sim 2 -\rho$, so that $K_{ab} \sim 2 J_aJ_b - \rho J_a J_b$.  

A few physical assumptions prove convenient  
to facilitate the formulation of  the quantum algorithm.

First, we restrict to the case in which dissipation is characterized 
by a single viscosity parameter \(\nu\), and we take an explicit expression
for the enthalpy so as to obtain a closed system.
Following the approach of~\cite{Sanavio_2024}, we assume a barothropic 
closure, in which the pressure depends only on the density and obeys the ideal-gas 
relation $P(\rho(\mathbf{r},t)) = \rho(\mathbf{r},t)\, c_s^2 \,.$
In this case, the enthalpy reduces to
\begin{equation}
\label{eq:enthalpy_perf_gas}
h(\rho) = \int_{\rho_0}^\rho d\rho'\,\frac{c_s^2}{\rho'} = c_s^2 \, \ln{\frac{\rho}{\rho_0}} \,.
\end{equation}
Furthermore, we consider the weakly compressible limit, assuming $\rho(\mathbf{r},t)\simeq \rho_0$. 
In this approximation, the enthalpy~\eqref{eq:enthalpy_perf_gas} can be linearized as
\begin{equation}
\label{eq:enthalpy_explicit_small_rho}
h(\rho) \simeq c_s^2 \left(\rho - \rho_0\right) + \mathcal{O}(\rho - \rho_0)^2,
\end{equation}
where $\rho_0$ is a reference density, which we set to unity for simplicity ($\rho_0 = 1$).


\section{Searching for a quantum algorithm: Carleman embedding}\label{sec:carleman_embedding}

Having discussed the derivation of SNS, we next proceed to the development of the
corresponding quantum algorithm. 
Although the SNS takes a quantum-like appearance, it contains several 
nonlinearities that pose significant challenges for a quantum implementation. 
First, the dissipative term, which depends on the Laplacian of 
the phase of the wavefunction. 
Being both non-polynomial and non-local, this term represents a major 
hurdle to the linearization of the SNS.  
\\
Additionally, in order to eliminate the quantum potential in the Madelung 
representation, this term is explicitly subtracted in the SNS via Eq.~\eqref{eq:potential_u},
introducing a further non-local nonlinearity.  

It is therefore convenient to turn back to a Madelung representation with $\rho$ and
$\theta$ as primary variables instead of the wavefunction itself.
We refer to this system as the Navier–Stokes equations in Hamilton–Jacobi form (NSHJ).  
\\
In this formulation, the quantum potential is removed and the viscous term is recast as a 
series of non-local polynomial nonlinearities. 
Although NSHJ contains several nonlinear terms, they are all 
second order, hence well suited to Carleman linearization. 
To prepare the system for the Carleman embedding, we first discretize the NSHJ equations by
means of a simple first-order Euler scheme for the time derivatives and centred finite differences for the spatial derivatives.  While this approach is known to be prone to numerical instabilities 
at both high and low Reynolds numbers~\cite{leveque2007finite, taylor_lewis_2009}, it is 
adequate for our purposes of assessing the performance of the Carleman 
embedding relative to the original NSHJ system.
More advanced discretizations may well lead to a better convergence of the Carleman 
procedure, an important research topic still awaiting for a systematic investigation. 
\\
For simplicity, we restrict the discussion to a two-dimensional setup, 
where $D_i$ denotes the finite difference operator along the $i-$th axis. 
The set of 2D discrete equations reads

{\small
     \begin{align}
        \label{eq:2D_continuity}
         &\hat{\rho} = \rho - \Delta t \left[
         (D_x\rho) v_x + (D_y \rho )v_y + \rho  (D_xv_x + D_yv_y) 
         \right] \, ;
         \\ \label{eq:2D_HJ}
         & \hat{\chi} = \chi + \Delta t \left[
          \nu ( D_xv_x + D_yv_y) - c_s^2\rho - \frac{1}{2}(v_x^2 + v_y^2)
         \right] \, ;
         \\    \label{eq:2D_Ax}
         & \hat{A_x} = A_x + \Delta t \left(
                \omega  v_y - \nu D_y \omega   
         \right) \, ;
         \\    \label{eq:2D_Ay}
         & \hat{A_y} = A_y + \Delta t \left(
              \nu  D_x\omega - \omega v_x 
         \right) \,;
     \end{align}
}
with the scalar vorticity
 $\omega = D_x A_y - D_y A_x$ and the hat symbol $\hat{\cdot}$ referring 
 to the time-evolved function, i.e., $\hat{\rho} = \rho(t+\Delta t)$. 

\subsection{Carleman Embedding}
The goal of Carleman approach is to recast the nonlinear system in Eqs.~\eqref{eq:2D_continuity}-~\eqref{eq:2D_Ay} as a 
formally infinite-dimensional linear one~\cite{Carleman}. 
This can be obtained by defining the first order Carleman state vector as
a collection of the following four fields:
\begin{equation}\label{eq:J_carleman}
    J_{\alpha} = 
    \begin{pmatrix}
        \rho \\\chi \\A_x \\A_y 
    \end{pmatrix} \,,\;\;\;\alpha=\{0,1,2,3\}
\end{equation}
where each component lives in two spatial dimension. 
Hence, on a grid with $G$ gridpoints , the first order Carleman vector
contains $4G$ variables.
The second order term $J^{(2)}_{\alpha \beta}(x,y) \equiv J_\alpha(x) \otimes J_\beta(y)$ 
originates from the product terms between fields, consisting of $16 \;G^2$ variables. 

With this notation, the starting system of equations takes the compact form 

\begin{equation}\label{eq:structure j}
    \hat{J}_\alpha (i) =  \sum_{i} A_{\alpha\beta}(ij) J_\beta(j) + \sum_{jk} B_{\alpha \beta \gamma}(ijk) J^{(2)}_{\beta \gamma}(jk) \, ,
\end{equation}
up to an irrelevant additive constant, which cancels out upon taking the gradient in the definition of the velocity~\eqref{eq:definition_v}. 
Here $i,j,k$ denote a flattened representation of the spatial indices and
the explicit expressions for the matrices $A$ and $B$ are given in the Supplementary Material~\ref{suppl:matrices}.
While the matrix $B$ acts on the full non-local tensor $J^{(2)}(\mathbf{r_1}, \mathbf{r_2})$, the relevant contributions are obtained by projecting onto its diagonal $\text{diag}(B_{\alpha\beta\gamma}J^{(2)}_{\beta\gamma})$; this enables to recover the right structure of Eqs.~\eqref{eq:2D_continuity}--~\eqref{eq:2D_Ay} (see Supplementary Information for details~\ref{suppl:matrices}).

The Carleman hierarchy is formally infinite, as each term is coupled to higher-order ones. 
For practical purposes, a closure is therefore required. 
In this work, we adopt a truncation closure: a finite Carleman order 
$N_C$ is chosen and all terms of order higher than $N_C$ are set to zero.
Better closures, based on linear dependencies on lower order Carleman levels
can certainly be devised, but they are beyond the scope of the present paper. 

\subsubsection{Second-order truncation}

We define the second order lifted state as:
\begin{equation}
\mathcal{J} =
\begin{pmatrix}
J_{1} \\
J_{2}
\end{pmatrix}
=
\begin{pmatrix}
J \\
J \otimes J
\end{pmatrix}.
\end{equation}

The evolution of the second-order state is obtained by 
the tensor product of the updated state~\eqref{eq:structure j} with itself:
\begin{align}
\hat{J}^{(2)} &= \hat{J} \otimes \hat{J} \\ \nonumber
&= (A J + B J^{\otimes 2}) \otimes (A J + B J^{\otimes 2}) \\ \nonumber
&= (A \otimes A) J^{\otimes 2} 
+ (A \otimes B + B \otimes A) J^{\otimes 3} 
+ (B \otimes B) J^{\otimes 4}.
\end{align}

In the second-order Carleman approximation, all terms of order three and higher are neglected:
\[
J^{\otimes 3}, \, J^{\otimes 4} \approx 0.
\]
The resulting truncated system reads
\begin{equation}
\label{eq:evol_eq}
\begin{pmatrix}
\hat{J}_1 \\
\hat{J}_{2}
\end{pmatrix}
=
\begin{pmatrix}
A & B \\
0 & A \otimes A
\end{pmatrix}
\begin{pmatrix}
J^{(1)} \\
J^{(2)}
\end{pmatrix},
\end{equation}
where the matrix \(A\) can be written as the sum of the identity and a term linear in $\Delta t$
\begin{equation}
A=\mathbb{I} + \Delta t \tilde{A},
\end{equation}
where $\tilde{A}$ incorporates the finite-difference operators in Eq.~\eqref{eq:2D_continuity}-~\eqref{eq:2D_Ay} and encodes the information needed to perform a single time step.

In this way, the tensor product $A \otimes A$ also contains quadratic 
terms in $\Delta t$, which can be neglected in the limit $\Delta t \to 0$, yielding the approximation
$\mathbb{I} \otimes \mathbb{I} + \Delta t \left( \tilde{A} \otimes \mathbb{I} + \mathbb{I} \otimes \tilde{A} \right)$. In the present work, however, we have chosen to retain these quadratic contributions.

\subsubsection{Tensor Network Representation}

\begin{figure}[tp]
    \centering
    \includegraphics[width=\linewidth]{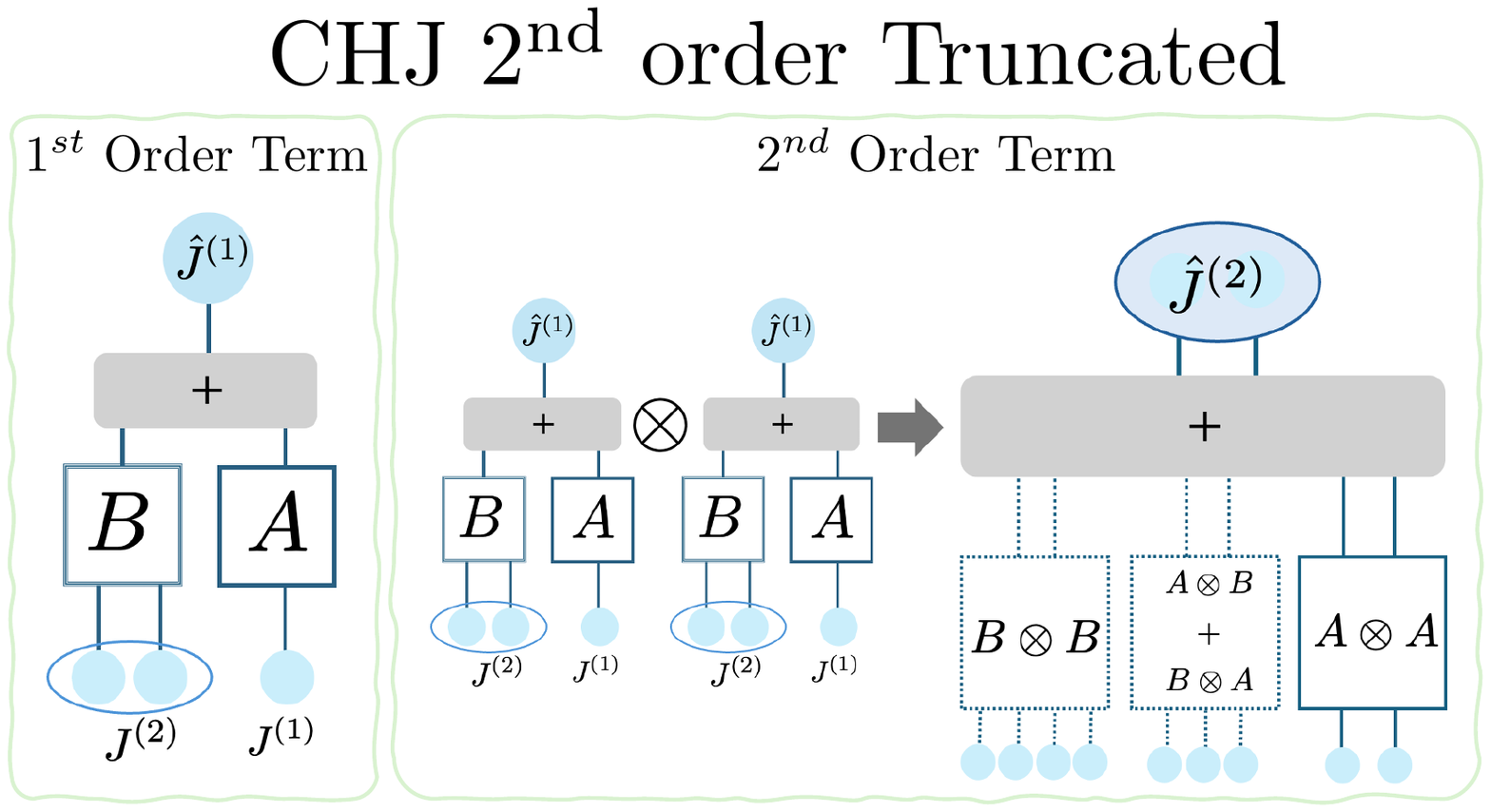}
    \caption{Tensor network representation of the second order NSHJ scheme. 
    On the left, the tensor representation of NSHJ; on the right, the product of two first NSHJ system gives the terms in the second order Carleman term. Choosing a truncation at second order, only the $A\otimes A$ term remains (solid line).}
    \label{fig:TN_higher_order}
\end{figure}

The generic order-$N_C$ expansion is obtained in a similar manner by taking the $N_C$-th power of Eq.~\eqref{eq:structure j}. While an explicit arithmetical formulation is 
possible, an intuitive approach based on a tensor network representation proves more convenient. 
In this representation, first-order state vectors $J_\alpha$ are depicted as circles, while 
higher-order states are shown as ovals; operators acting on these states are represented 
as squared boxes, and lines denote free or contracted tensor indices. 
Following this convention, the NSHJ system can be visualized as in Fig.~\ref{fig:TN_higher_order}. 
Combining $m$ such "Carleman diagrams" yields the evolution of the $m$-th term, with 
truncation applied by neglecting all terms of order higher than $N_C$. An example is reported in Fig.~\ref{fig:TN_higher_order}.
This representation also brings a significant practical advantage as it allows the 
evolution of the Carleman system to be expressed using a tensor network structure, which 
can be exploited to significantly reduce the computational complexity. 
All numerical simulations in this work were performed efficiently thanks to this strategy.
Particularly, $N_c=4$ simulations would have not been possible at all without the tensor
network representation, which requires  $\mathcal{O}(10^2)$ Gigabyte 
instead of $10^5$, as shown in~\ref{suppl:TN_representation}.
More details on such procedure are given in the Supplement Material~\ref{suppl:TN_representation}.

\section{The quantum algorithm}
\label{sec:quant_alg}

    In this section we describe how to formulate a quantum algorithm to perform the evolution step described in Eq.~\eqref{eq:evol_eq} above, leaving a more detailed implementation to future work.
    As a first step we take the number of gridpoints $G$ to be a power of $2$ in order to curb
    the qubit encoding overhead ~\footnote{Of course this requirement can be relaxed at the expense of suitably padding the resulting vectors and matrices with zeros.}. In particular, we assume 
    $4G=2^n$ so that $n$ qubits are required to store the $J^{(1)}$ vector 
    and $2n$ qubits are needed instead for $J^{(2)}$. 
    In this scheme, the quantum state at any given time $t$ encodes the solution vectors as
    \begin{equation}\label{eq:wavefunction_psi}
    \rvert\Psi(t)\rangle = \rvert J^{(1)}(t)\rangle\otimes\rvert J^{(2)}_\alpha(t)\rangle\otimes\rvert J^{(2)}_\beta(t)\rangle\;,
    \end{equation}
    where we used the fact that the tensor factorization of the $J^{(2)}_{\alpha\beta}$ vector is maintained during the evolution step. With this encoding, the evolution step presented in Eq.~\eqref{eq:evol_eq} takes the following form
    \begin{align}\label{eq:quantum_algorithm}
    \rvert\Psi(t+\Delta t)\rangle 
    & 
    = \left(\mathbb{I}\otimes A\otimes A\right)\left(A\otimes\mathbb{I}\otimes\mathbb{I}+\overline{B}\right)\rvert\Psi(t)\rangle
    \\ & \nonumber
    =M_2M_1\rvert\Psi(t)\rangle\;,
    \end{align}
        where the $2^n\times2^n$ matrix $A$ acts non trivially only on one register of size $n$ at a time, while $\overline{B}$ is a matrix acting on the full space with the top right $2^n\times4^n$ rectangular sub-matrix equal to $B$ and the rest zero. In this formulation we first perform the update on the $J^{(1)}$ component, described by the matrix $M_1$, followed by the update of the $J^{(2)}$ component using the $M_2$ matrix.
    
    Since the matrices involved are not unitary, in order to implement the update step on a 
    quantum computer we need to encode them within a unitary operator. 
    A standard procedure is the so-called block-encoding~\cite{camps2024, low2019,gilyen2019quantum} 
    that uses a given number of ancilla qubits $m$ to embed a generic $2^q\times2^q$ matrix $M$ into 
    a unitary operator acting on the full register of $(m+q)$ qubits, with the following block structure
    \begin{equation}
        U_M=
        \begin{pmatrix}
            M/\alpha_M & * \\
            * & *
        \end{pmatrix},
        \label{eq:block_encoding}
    \end{equation}
    In the above, the rescaling factor $\alpha_M\geq\|M\|$ guarantees the realizability
    of the embedding. 
    The actual value of $\alpha_M$ depends on the specific implementation used to implement the block encoding.

    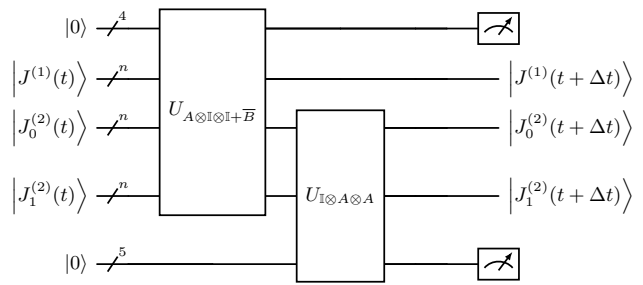
\begin{figure}[tp]
    \resizebox{\linewidth}{!}{
        \centering
        \begin{quantikz}
            \lstick{$\ket{0}$}          & \qwbundle{4} & \gate[4]{U_{A\otimes\mathbb{I}\otimes\mathbb{I}+\overline{B}}} & \qw & \qw & \qw & \meter{} \\
            \lstick{$\ket{J^{(1)}(t)}$} & \qwbundle{n} &       & \qw & \qw & \qw & \rstick{$\ket{J^{(1)}(t+\Delta t)}$} \\
            \lstick{$\ket{J^{(2)}_0(t)}$} &\qwbundle{n}&       &\gate[3]{U_{\mathbb{I}\otimes A\otimes A}} & \qw & \qw & \rstick{$\ket{J^{(2)}_0(t+\Delta t)}$}\\
            \lstick{$\ket{J^{(2)}_1(t)}$} &\qwbundle{n}&       &                                           & \qw & \qw & \rstick{$\ket{J^{(2)}_1(t+\Delta t)}$}\\
            \lstick{$\ket{0}$}            &\qwbundle{5}& \qw   &                                           & \qw & \qw & \meter{} \\
        \end{quantikz}
        }
        \caption{Quantum circuit scheme representing the block-encoding of the evolution step from Eq.~\eqref{eq:quantum_algorithm}. We also explicitly show the $4$ ancilla qubits used for the block encoding of the $M_1$ operator on the top and the $5$ qubits for $M_2$ on the bottom.}
        \label{fig:dec_circ}
    \end{figure}

   A  possible approach to implement these matrices on a quantum computer is to exploit their sparsity. 
   This can be achieved by combining a $SELECT$ oracle, which identifies and applies the 
   nonzero block terms, combined with an efficient block encoding of the individual blocks, as detailed in the Supplemental material~\ref{suppl:matrices}. The construction we use also relies on the linear combination of unitaries~\cite{Childs:2012gwh} expansion to combine together different sparse matrices that are encoded separately. 
   We encode the operators $M_1$ and $M_2$ with separate block encodings and apply them to the system in 
   sequential succession.

    However, since the underlying evolution matrix is generally nonunitary, the implementation succeeds only probabilistically: the desired operation is realized on the subspace where all ancilla qubits are in the $\rvert0\rangle$ state. 
    The corresponding success probability depends on sub-normalization constants and the norm of the
    wavefunction in Eq.~\eqref{eq:wavefunction_psi}.
    With our implementation we find (see Eq.~\eqref{eq:succ_prob_app} in the Supplemental material~\ref{suppl:matrices}
    \begin{align}   
    p_s(t \to t + \Delta t) 
    & = \frac{ \left\| \Psi (t+\Delta t)\right\|^2}{\alpha^2_{M_1}\alpha^2_{M_2}}
    \\ & \nonumber
    =\left\| \Psi (t+\Delta t)\right\|^2\left(1+O\left(\frac{\Delta t}{\Delta x^2}\right)\right)\;,
    \end{align}
    where the asymptotic form is valid only in the limit $\Delta t/\Delta x^2\ll1$. 
    With the parameters used in this work ($\Delta t=0.01$, $\Delta x=2\pi/32 \sim 0.2$, $c_s^2=1/3$ and $\nu=1/6$), this 
    condition is not met, resulting in $\alpha^2_{M_1}\alpha^2_{M_2}\approx 10^5$. 
    The success probability is therefore comparable to the one obtained with the Carleman Lattice Boltzmann 
    using the D2Q9 model \cite{Sanavio_25_PoF} which has a similar sparsity pattern. 
    By reducing further the time-step to $\Delta t=0.001$ we find instead $\alpha^2_{M_1}\alpha^2_{M_2}\approx 60$,
    which is nonetheless less efficient since it requires ten times more steps. 
    In this regime of large $\Delta t/\Delta x^2$ a different block encoding, or a different mapping to qubits, might prove more efficient.

    In order for this probability to remain sizable, the norm of the Carleman vector should 
    decay as slowly as possible. In the next subsection we provide numerical evidence that 
    this is indeed a reasonable expectation.

    A more complete resource estimate of the cost of implementing the present quantum algorithm 
    is beyond the scope of the present article, and will make the object of future research.

    \subsection{Amplitude Decay}
    In order to ensure the proper decay of the Carleman vector $\mathcal{J} = (J^{(1)}, J^{(2)})^T$, it is essential to add a constant factor $K = (0, c_s^2, 0, 0)^T$ to the Carleman system. In principle, one could add any arbitrary quantity 
   to $\chi$, which is not a direct observable and appears only through its gradient. 
   In other words, it is gauge invariant and a gauge transformation should be fixed in order to recover appropriate amplitude decay of the Carleman vector. Without this constant factor, the norm would grow, preventing the development of the quantum algorithm.

\begin{figure}
        \centering
        \includegraphics[width=\linewidth]{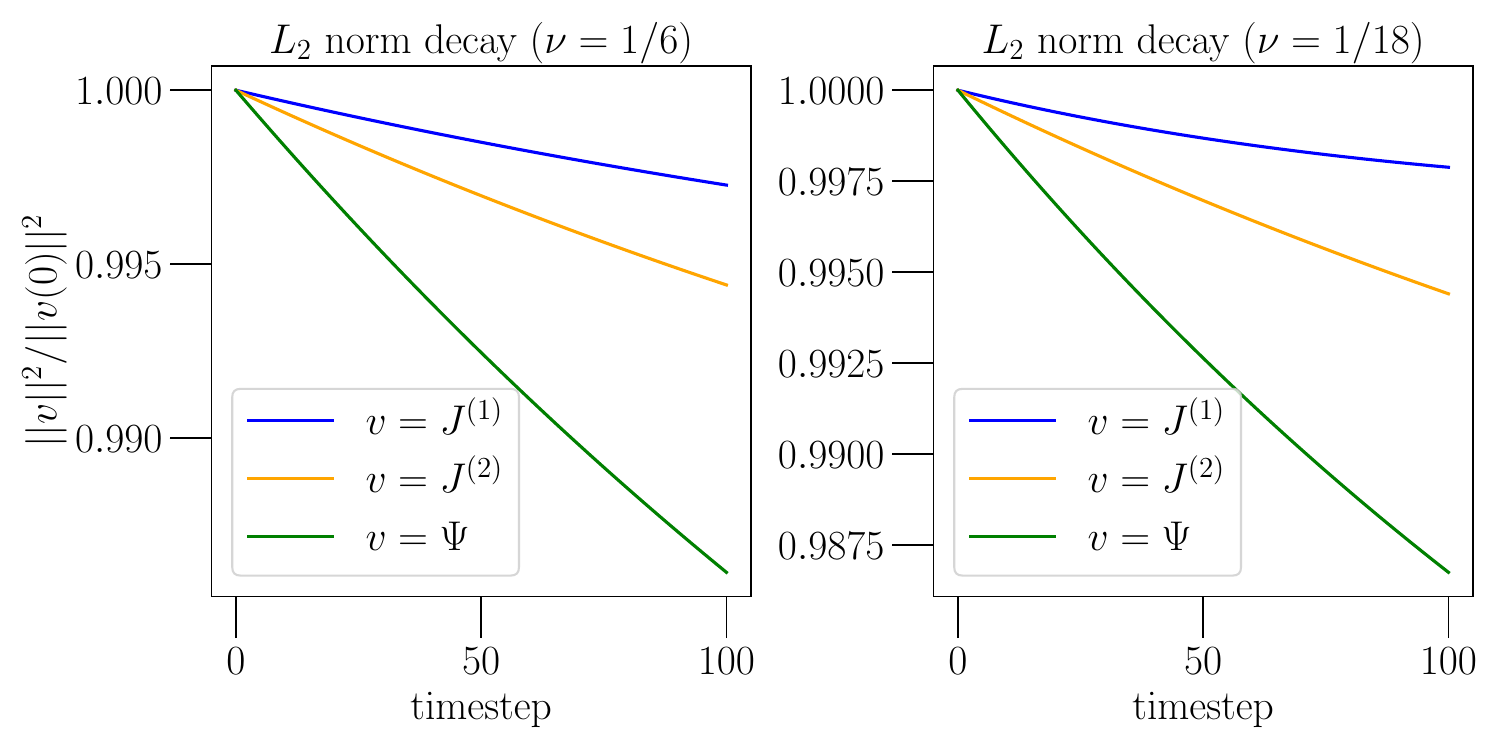}
        \caption{Decay rate of the $L_2$ norm for $J^{(1)}, J^{(2)}$ and $\Psi$ in Eq.~\ref{eq:wavefunction_psi} for different value of the viscosity coefficient $\nu$. The case in example refers to the initial condition in Eq.~\ref{eq:initial_conditions}.
        }
        \label{fig:norm_carleman}
    \end{figure}

    We examined the decay of the Carleman vector components \(J^{(1)}\) and \(J^{(2)}\), along with the wavefunction \(\Psi\) as defined in Eq.~\eqref{eq:wavefunction_psi}, for different values of \(\nu\), corresponding to different Reynolds numbers. We reported the results in Fig.~\eqref{fig:norm_carleman} where the decay rate of \(\Psi\) is in the order of $10^{-2}$ over $100$ timesteps. 
All quantities are normalized to their initial values to allow a clear comparison.
    
\section{Numerical Simulations}\label{sec:numerical_simulations}

In this section, we investigate the approximation of the NSHJ system using the Carleman approach, with the goal of comparing it with previous formulations, in particular the CLB~\cite{sanavio_CLB}  and Carleman Navier-Stokes (CNS) formulations. To facilitate a direct comparison, we adopt the same physical system and initial conditions as in~\cite{Sanavio_2024}.

We consider a system on a periodic square grid with $G = 32 \times 32$ points and 
box size $L_{\text{box}} = 2\pi$. The initial condition is chosen to represent a Kolmogorov-like flow:
\begin{align}\label{eq:initial_conditions}
    &\rho = 1, \hspace{2.3cm} \chi = 0, 
    \\&\nonumber
    A_x = u_x \cos(k_x y), \quad A_y = {u_y} \cos(k_y x),
\end{align}
with $k_x=k_y=1$, $u_x=u_y=0.1$ and $c_s^2 = 1/3$. 
In this configuration the Mach number is approximately 
$Ma \sim 0.17$, in line with the assumption of a weakly compressible regime.
The corresponding Reynolds number is given by:
\begin{equation}\label{eq:reynolds}
    Re = \frac{||u_0|| L_{\text{box}}}{\nu}, 
\end{equation}
where $||u_0||$ denotes the $L_2$ norm of the velocity field. 
The system exhibits a decay dynamics characterized by the 
dissipative timescale 
\begin{equation}
  T = \frac{1}{k^2 \nu}\,,
\end{equation}
where \(k\) denotes the dominant wavenumber in Eq.~\eqref{eq:initial_conditions}.

To quantify the discrepancy between the flow fields generated by the Carleman approach and 
the NSHJ solution, we introduce a global error metric. 
We employ a normalized $L_2$ norm, which avoids spurious amplification of 
relative errors in regions where the field amplitude becomes vanishingly small. 
Specifically, we define
\begin{equation}
    \label{eq:L2_error}
    \frac{\Delta J_\alpha}{J_\alpha} = \frac{\lVert {J_\alpha}_{\mathrm{C}} - J_{\mathrm{NSHJ}} \rVert_{L_2}}{\lVert {J_\alpha}_{\mathrm{NSHJ}} \rVert_{L_2}} \, .
\end{equation}
Here $J{_\alpha}$ is the Carleman state as defined in Eq.~\eqref{eq:J_carleman}, with the subscript 
C referring to the linearized system and NSHJ to the reference solution of the NSHJ equations.

\begin{figure*}
    \centering
    \includegraphics[width=.9\linewidth]{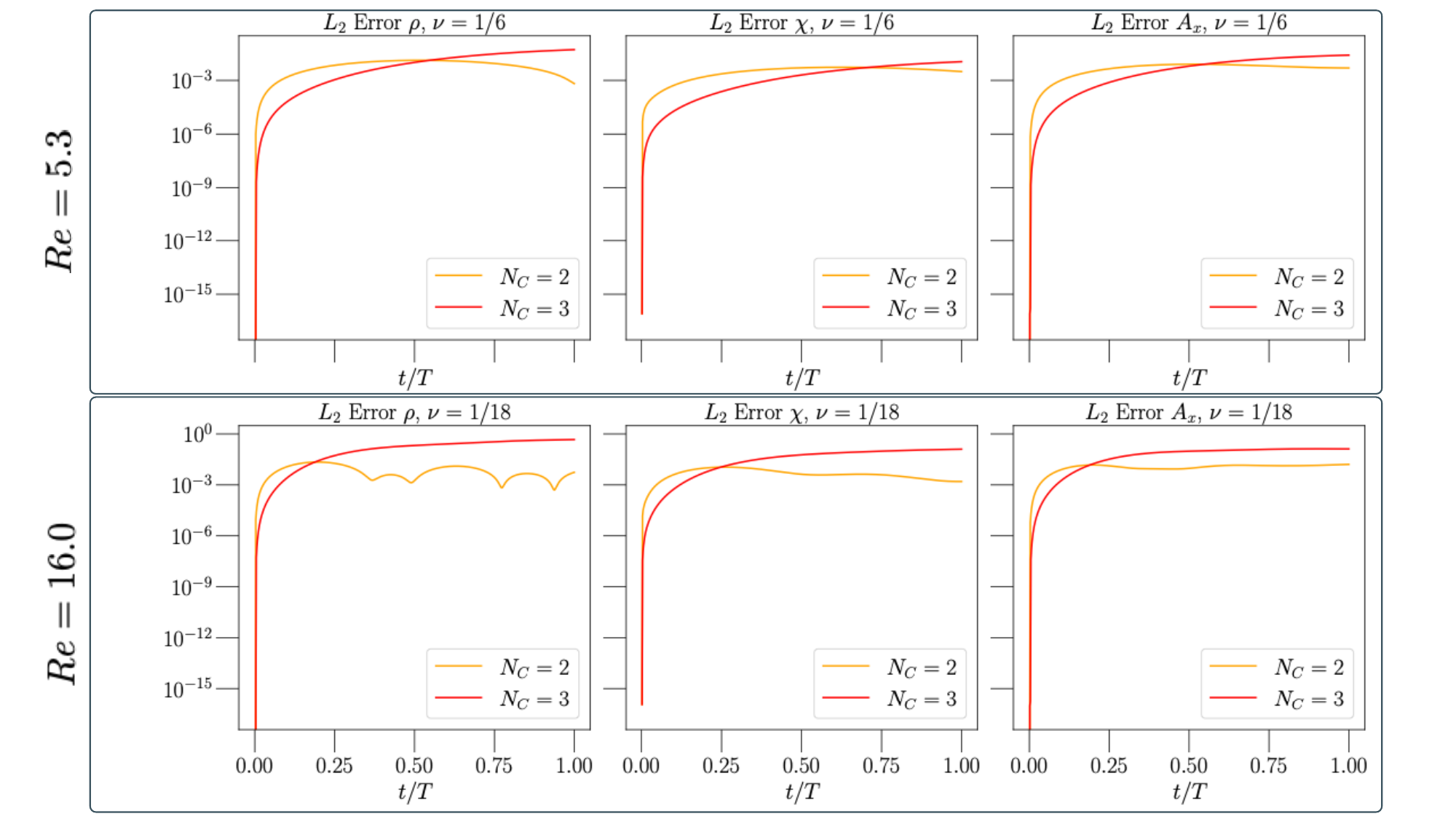}
    \caption{Global relative error of the NSHJ fields ($\rho, \chi, A_x$) for different Carleman truncation order $N_C$. The initial condition in Eq.~\eqref{eq:initial_conditions} is evolved for 600 timesteps on a $32\times 32$ grid. In the first row, the results are obtained with $\nu=1/6$; in the second row with $\nu = 1/18$. 
    From the figure, it can be observed that increasing $N_C$ enhances the accuracy up to a 
    crossover time $t_{\textit{cross}} < T$, whose value depends on the Reynolds number. 
    }
    \label{fig:fields_comparison}
\end{figure*}

\subsection{Comparing Carleman fields}
As a first test, we examined the ability of the Carleman representation (CHJ) to reproduce the NSHJ fields. 
In particular, we focus on the way that the Carleman variable $J$ captures $\rho$, $\chi$, $A_x$, and $A_y$. 
While these quantities are not the primary physical observables, they are the fields 
that CHJ is expected to reproduce, according to Eqs.~\eqref{eq:2D_continuity}-~\eqref{eq:2D_Ay}.

We consider the system defined by Eq.~\eqref{eq:initial_conditions} for $\nu = 1/6$ and $\nu = 1/18$, corresponding to Reynolds numbers $Re \approx 5.33$ and $Re \approx 16$, respectively. Figure~\ref{fig:fields_comparison} shows the evolution of the system up to one dissipative time $T$, computed over $600$ time-steps for both cases. Using the full Carleman matrix, the simulation would have required $\mathcal{O}(100) GB$ of memory; thanks to the tensor network approach presented in the Supplemental Material~\ref{suppl:TN_representation},  the cost is dramatically reduced 
to about $\mathcal{O}(10^{-2}) GB$.

The results show that the truncation order affects accuracy over different timescales. 
At early times, the third-order truncation ($N_C=3$) performs better than the 
second-order ($N_C=2$), reflecting the improved linearization. 
At longer times, however, the second-order scheme becomes more 
accurate, consistently with the slower relaxation of the system. 
\\
Specifically, we observe that the crossover timescale \(t_{\text{cross}}\) marks the 
moment when the error of the third-order CHJ approximation becomes less accurate than the second-order one. 
This crossover timescale depends on the flow regime: with \(\nu = 1/18\) (\(Re \sim 16\)) it occurs 
around \(T/4\), whereas for \(\nu = 1/6\) it shifts closer to \(T/2\). Importantly, in both cases, the second-order Carleman 
truncation maintains relative errors below \(10^{-2}\), despite the lack of a systematic convergence.

Consequently, the Carleman linearization provides a reasonable approximation only 
within a finite temporal window, whose extent is controlled by the Reynolds number. 
This behaviour is consistent with established results in Carleman theory 
and previous studies in the literature, as discussed in~\cite{forets2017}.
\\
These findings also confirm the validity of the Carleman approach at the level 
of native Carleman fields $J_{\alpha}$. 
For physically purposes, however, the velocity or momentum fields must be 
used, as they represent the quantities directly relevant to the system’s dynamics.

\subsection{Short Time Behaviour}

Recalling the expression for the velocity field in Eq.~\eqref{eq:definition_v}, 
we define the momentum as $\boldsymbol{J}(\boldsymbol{r},t) = \rho(\boldsymbol{r},t) \boldsymbol{v}(\boldsymbol{r}, t)$.
With this definition, it is possible to set a direct comparison between 
the CHJ results and the CNS ones presented in~\cite{Sanavio_2024}, based
on the conservative formulation of the Navier-Stokes equations. 

The system in Eq.~\eqref{eq:initial_conditions} is evolved up 
to $t=T$ on a $32\times 32$ grid, with different viscosity coefficients: $\nu=1/6$ and $\nu=1/18$. 
For this comparison, we ran the simulation up to $T/4$, corresponding to a 
total of $150$ time steps in both cases, where we have set 
$\Delta t = 0.01$ for $\nu=1/6$, ensuring that the simulation remains 
comparable to the CNS results.

The present simulations were performed on a laptop with an M2 processor, which was only
possible thanks to the tensor network strategy outlined in the Supplemental 
Material~\ref{suppl:TN_representation}, given that the 
fourth order Carleman matrix would have required $\mathcal{O}(10^5)GB$.

We did not extend the simulation beyond $T/4$ for the reasons discussed 
previously: at short time-scales, where $t\ll T$ convergence improves with higher 
$N_C$ only up until a time limit $t_{\text{cross}}$. 
Short time simulations are therefore sufficient to perform the 
analysis of CHJ up to the fourth order and compare the results with CNS.

\begin{figure*}
    \centering
    \includegraphics[width=.9\linewidth]{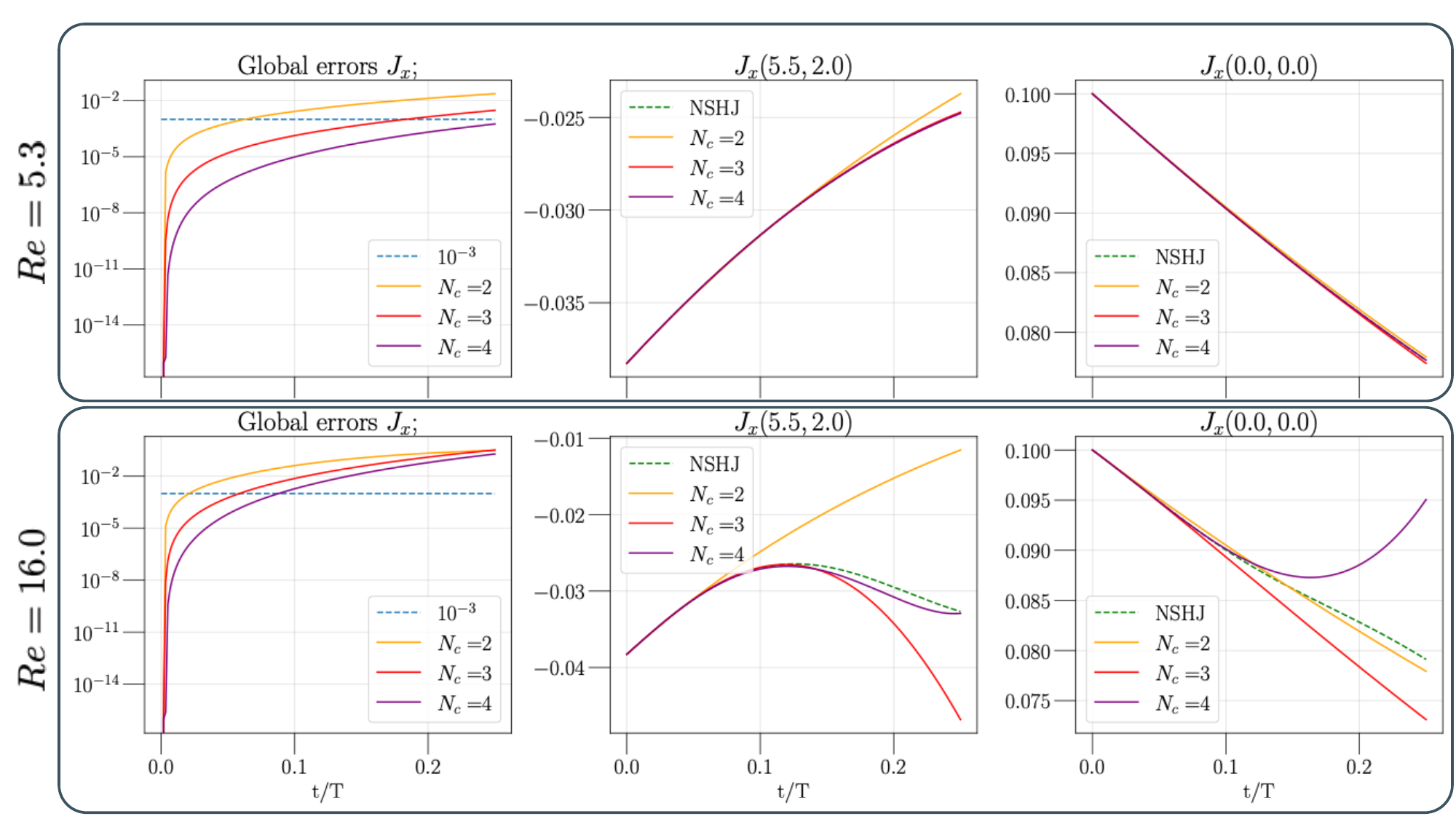}
    \caption{
    Comparison of the momentum field $J_x$ obtained from the initial conditions in Eq.~\eqref{eq:initial_conditions} with NSHJ and its Carleman version, for $\nu=1/6$ (first row) and $\nu=1/18$ (second row). Both simulations run for 150 timesteps total on a $32\times 32$ grid. 
    The first two columns show the evolution of $J_x$ at two distinct grid points for different 
    Carleman truncation orders ($N_C$). 
    The first column reports the evolution of the global relative error, with the green dotted 
    line representing the $10^{-3}$ error threshold. 
    The figure illustrates the short term behaviour, where increasing $N_C$, improves the accuracy 
    and convergence time. Moreover, it shows that CHJ is sensitive to the Reynolds number. 
    The results in the first row show that CHJ4 keeps the error below $10^{-3}$ all along the evolution. 
    In the second row, with higher Reynolds number, CHJ4 crosses the $10^{-3}$ 
    threshold around $T/10$, indicating a dependence on the Reynolds number.
    }
    \label{fig:carleman4_k1}
\end{figure*}

In Fig.~\ref{fig:carleman4_k1}, we show the evolution of the momentum field
 along the $x$-axis for two different Reynolds numbers. 
 We present the time evolution of the global $L_2$ error of  $J_x$  and the value of the field 
 at two spatial locations: $(0,0)$ corresponding to one of the maximum values 
 of the initial condition and $(5.5, 2)$ where initial amplitude close to zero .

\begin{figure*}
    \centering
    \includegraphics[width=.9\linewidth]{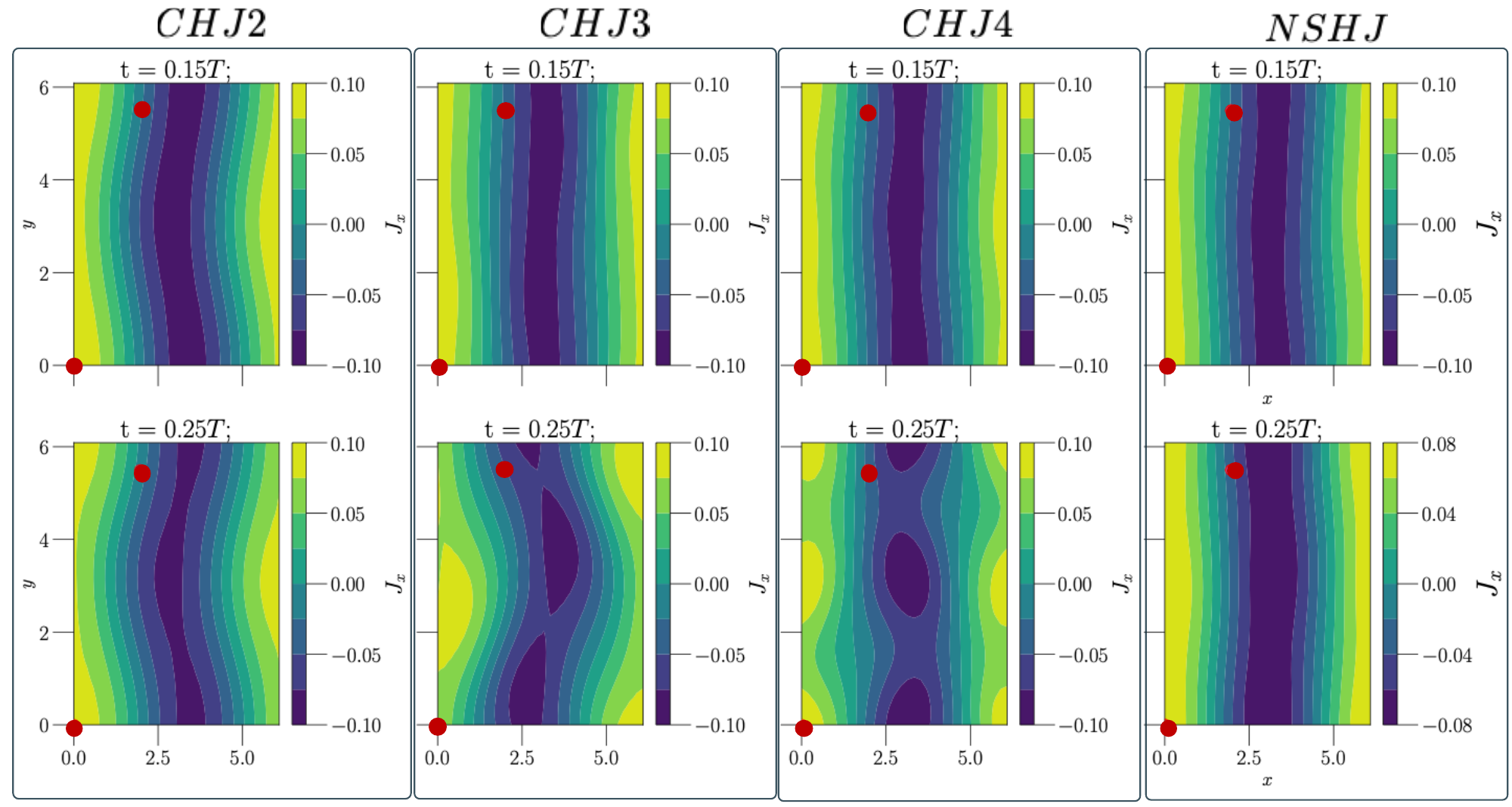}
    \caption{Color map of the momentum field $J_x(x,y)$ at different times for CHJ and NSHJ. 
    The figure shows the evolution of the initial condition in 
    Eq.~\eqref{eq:initial_conditions} on a $32 \times 32$ grid with $\nu = 1/18$. 
    The red dots mark the points where the time evolution of $J_x$ is monitored throughout the simulation.  
    Each column corresponds to a different Carleman truncation order $N_C$, with the rightmost 
    one showing the reference NSHJ solution. The first and second rows report the field at $t = 0.15T$ 
    and  $t = T/4$, respectively. Overall, the figure provides a visual counterpart to 
    the error analysis presented in Fig.~\ref{fig:carleman4_k1}.
    }
    \label{fig:field_comparison}
\end{figure*}

As already observed earlier on, the global error decreases with increasing 
truncation order within the time window $ t < t_{\text{cross}} $, with 
$ t_{\text{cross}} $ depending on both the truncation order and the Reynolds number.  
The first row shows that, for $ \nu = 1/6 $, the CHJ4 approximation reproduces the 
reference solution with good accuracy, keeping the relative error below $10^{-3}$ over the entire evolution. 
In fact, both CHJ4 and CHJ3 closely follow the reference curve, with no visible deviations. 
In this case, the crossover time $ t_{\text{cross}} $ exceeds the observation horizon $ T/4 $.  
As the Reynolds number increases, the Carleman approximations gradually depart 
from the reference solution, indicating a mounting influence of inertial effects on the CHJ dynamics.  

This behaviour is further illustrated in the second row of Fig.~\ref{fig:carleman4_k1}, where higher-order truncations remain consistent with the reference solution for longer times. 
This is reflected both in the error values shown in the first column and in the evolution 
of $ J_x(\mathbf{r}) $, with CHJ4 maintaining agreement over a wider time range.

\begin{figure*}
    \centering
    \includegraphics[width=.9\linewidth]{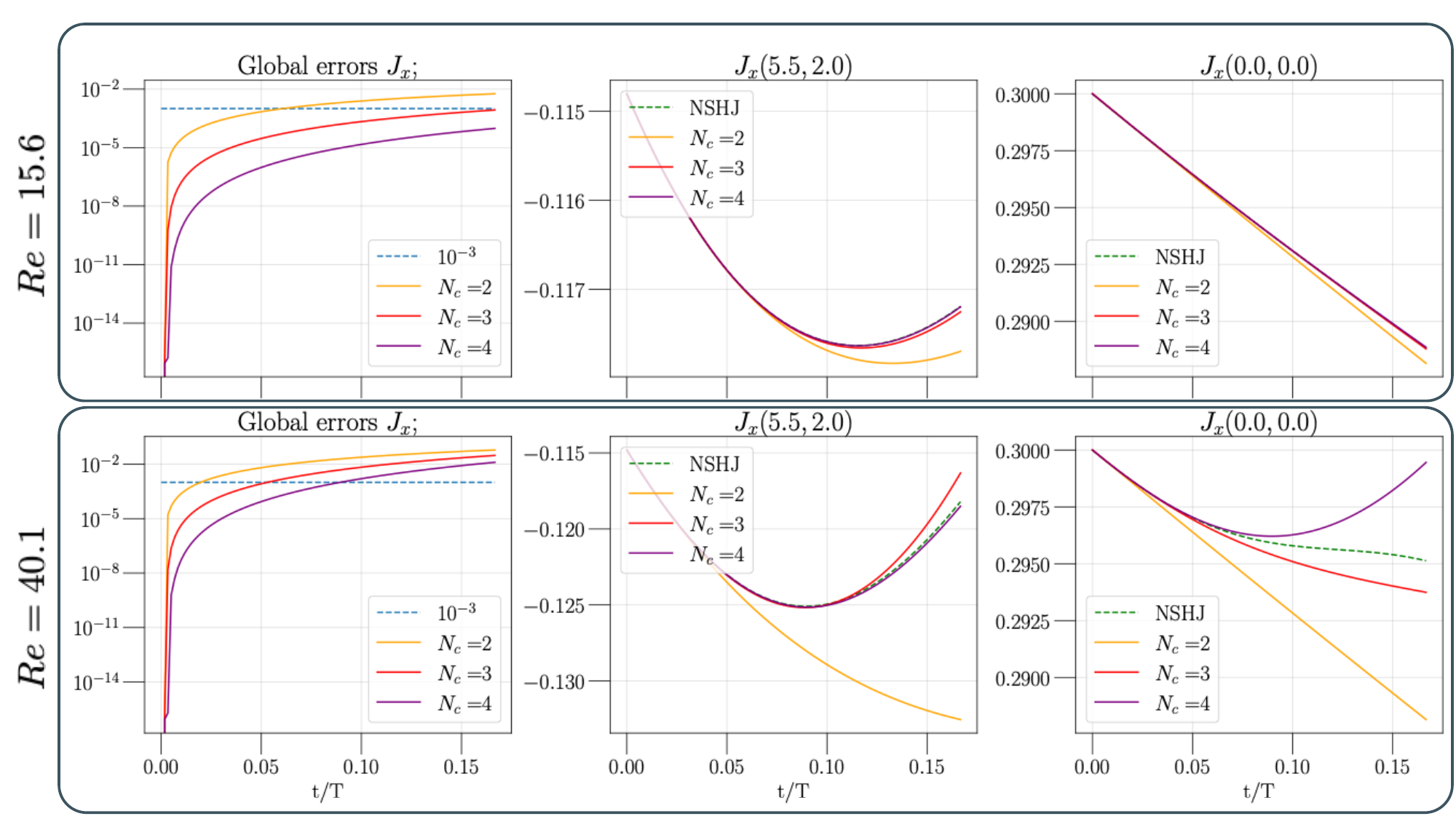}
    \caption{
    Comparison of the momentum field $J_x$ obtained with NSHJ and its Carleman versions, for $\nu=1/6$ (first row) and $\nu=1/18$ (second row). In both simulations the initial condition in Eq.~\eqref{eq:initial_conditions} with $(k_x, k_y) = (1,4)$ and $(u_x,u_y) = (0.3, 0.2),$ is evolved for 100 timesteps on a $32\times 32$ grid. 
    The first two columns show the evolution of $J_x$ at two distinct grid points for different Carleman truncation orders ($N_C$). 
    The first column reports the evolution of the global relative error, with 
    the green dotted line representing the $10^{-3}$ error threshold. 
    The figure illustrates the accuracy of CHJ in the regime of moderate Reynolds numbers for evolutions up to a fraction of the dynamical time $T$.
    }
    \label{fig:carleman4_k4}
\end{figure*}

Unlike the behaviour observed for the fields in Fig.~\ref{fig:field_comparison}, where the second-order CHJ (CHJ2) keeps the relative error below $10^{-2}$, here the error remains noticeably higher for second-order Carleman. This is expected, since computing the velocity field involves finite differences of $\chi$ and the addition of the field $\mathbf{A}$, which is a source of numerical inaccuracy.  

Fig.~\ref{fig:field_comparison} reports a visualization of the momentum field $J_x$ at $Re\approx 16$, obtained with different Carleman truncation orders at two distinct time-frames. At $t=0.15T$, there are no notable differences between CHJ4 and the reference NSHJ solution, showing excellent agreement. 
After 150 timesteps, corresponding to $t = T/4$, all Carleman approximations begin 
to drift visibly away from the exact solution. 
For the sake of clarity, we highlight the two spatial probes used to track the evolution of $J_x$ in Fig.~\ref{fig:carleman4_k1} as red circles.  

For comparison, the CNS approach in~\cite{Sanavio_2024} converged to the exact solution only up to $\mathcal{O}(10)$ timesteps for $\nu=1/6$ with Carleman truncation below $N_C = 4$. 
Using CHJ with fourth-order Carleman, we observe relative errors around $10^{-4}$, keeping 
a close agreement with the exact solution for over $\mathcal{O}(150)$ timesteps. 
This shows that the HJ reformulation of the NSE significantly improves over the 
accuracy of the Carleman embedding as applied to the Navier-Stokes equations.  

To further test the CHJ approach, we adopt the initial conditions from~\cite{Sanavio_2024}, corresponding to the Kolmogorov-like flow in Eq.~\eqref{eq:initial_conditions}, with $u_x = 0.3$, $u_y = 0.2$, and asymmetric wavenumbers $k_x = 1$, $k_y = 4$.  
The system is evolved up to $T/6$  (100 timesteps)  for both $\nu=1/6$ and $\nu=1/18$, corresponding to Reynolds numbers of approximately $14$ and $41$. 
The results in Fig.~\ref{fig:carleman4_k4} confirm that for moderate 
Reynolds numbers and $\nu=1/6$  CHJ correctly reproduces the target behaviour, with relative errors comparable to the ones obtained with CLB in~\cite{sanavio_CLB}.  
Increasing the Reynolds number results in a reduced coherence time with the exact solution. 
This is shown explicitly in the second column of Fig.~\ref{fig:carleman4_k4}, where 
the solution for $\nu=1/6$ $CHJ4$ overlaps with $NSHJ$, with $CHJ3$ departing from the exact solution 
before $t=0.15T$. In the case $\nu=1/18$ this is more evident, with $CHJ4$ slowly departing 
from the right solution at later times than the lower order approximations.

\begin{figure*}
    \centering
    \includegraphics[width=.9\linewidth]{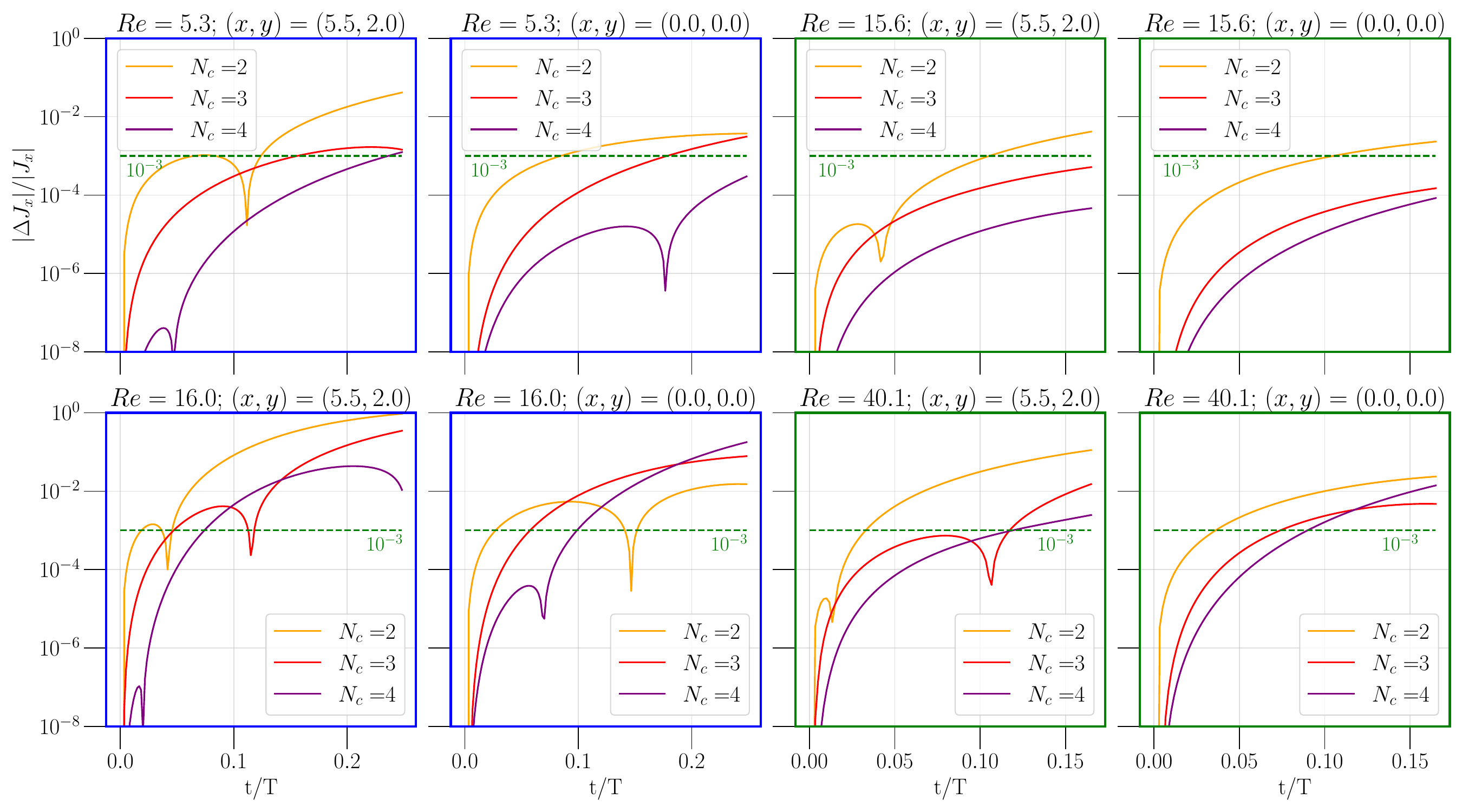}
    \caption{Evolution of relative errors $|\Delta J_x(\mathbf{r})| / |J_x(\mathbf{r})|$ in different points of the grid. 
    On the left, the blue-framed plots, report the local error for the points presented in Fig.~\ref{fig:carleman4_k1}. 
    On the right side of the image, green-framed, are presented the relative errors of the points shown in 
    Fig.~\ref{fig:carleman4_k4}. The cusps originate from occasional super-convergence of the Carleman approximation (inflexion points).
    }
    \label{fig:short_time_error_comp}
\end{figure*}

To support this remark, Fig.~\ref{fig:short_time_error_comp} shows the local relative error $ |\Delta J_x(\mathbf{r})| / |J_x(\mathbf{r})| $ for the points marked in Fig.~\ref{fig:field_comparison} and shown both in Fig.~\ref{fig:carleman4_k1} and Fig.~\ref{fig:carleman4_k4}, where the 
accuracy threshold, $\Delta x^2 \approx 10^{-3}$, is shown for reference. 
Even in this local comparison, CHJ4 maintains excellent agreement with 
NSHJ for $\nu=1/6$, despite the different Reynolds numbers. 
The presence of the cusp in the errors is interpreted as an occasional 
super-convergence event, which is rapidly absorbed to regain the standard error trend. 

\subsection{Long Time behavior}

The general spirit of Carleman linearization as a multiple timescale expansion
procedure, is to expand the convergence time horizon at increasing order of truncation. 
Once this time horizon is passed, higher order approximations 
usually depart from exact solution more rapidly than the low order ones. 
Indeed, it is observed that the lowest order case $N_c=2$ proves capable of 
recovering the long-term time behaviour to a reasonable accuracy.
This opens up an interesting connection with Tauberian theorems describing the
time-asymptotic behaviour of dynamical systems, which  may prove useful  
to capture statistical steady states rather than the detailed overall dynamics~\cite{belmabrouk2026}

To investigate this aspect  we analyse the consistency of the method for 
evolution times beyond the dissipative scale $T$. 
In particular, we focus on the possibility of recovering the stationary behaviour 
of the system using the lowest CHJ2 truncation.  

As shown in Fig.~\ref{fig:fields_comparison}, in this time regime the 
second-order Carleman expansion provides the most reliable approximation, aside 
from much higher-order truncations of order 20 or more. 
As observed in the previous section, increasing the truncation order 
$N_C$ extends the convergence time to the exact solution. 
For instance, for $\nu=1/6$ one can reach coherence times of $\mathcal{O}(100)$ timesteps.  
\\
However, high-order truncations of order $20$ or more are totally impractical. 
Therefore, in order to study the stationary behaviour of the initial conditions, we 
restrict our analysis to the second-order Carleman expansion. 
This choice balances accuracy and computational feasibility, while still capturing the 
essential long-term dynamics of the system.

\begin{figure*}
    \centering
    \includegraphics[width=.9\linewidth]{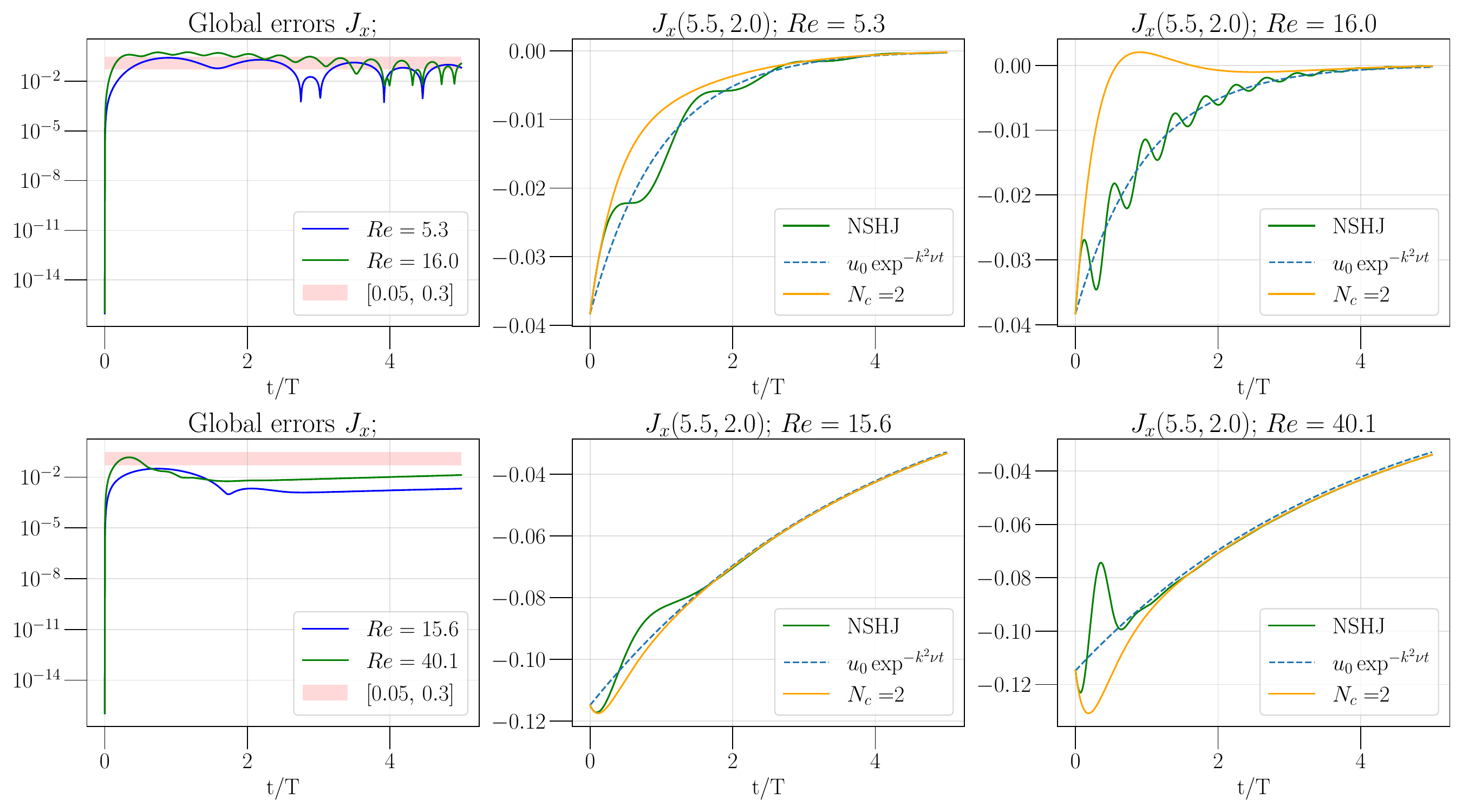}
    \caption{
    Comparison of the momentum field $J_x$ obtained evolving the initial condition in Eq.~\eqref{eq:initial_conditions} with NSHJ and its CHJ truncated at second order for different Reynolds numbers; both simulations run for 2400 timesteps on a $128\times 128$ grid. 
    Results in the first row are obtained with $u_x = u_y = 0.1$ and $k_x = k_y=1$, while in the second row $(u_x, u_y) = (0.3,0.2)$ and $(k_x, k_y) = (1, 4)$.
    The first column from the left reports the evolution of the global relative $L_2$ error for the two different Reynolds, with the red band indicating regions of high relative error.
    The center and left column show the evolution of $J_x$ on a gridpoint with initial value close to zero for  $\nu=1/6$ and $\nu=1/18$ respectively. 
    In both cases, the second-order truncation correctly captures the stationary state.
    }
\label{fig:Carleman_2_long}
\end{figure*}

The initial conditions defined in Eq.~\eqref{eq:initial_conditions} are evolved over  
 $5T$ with $\nu=1/6$ and $\nu=1/18$ on a $128 \times 128$ spatial grid. 
 After rescaling the dynamical time accordingly, both simulations are run for 
 a total of  $3000$ timesteps. 
 The errors and the evolution of $J_x(5.5,2)$, a point with small initial value, are 
 reported in the first row of Fig.~\ref{fig:Carleman_2_long}. 
 The second row shows the results obtained with the asymmetrical initial condition 
 $u_x = 0.3, u_y = 0.2$ and $k_x = 1, k_y = 4$.

In both cases, CHJ2 does not perfectly reproduce the exact NSHJ solution. 
However, it correctly captures the overall decay trend and stabilizes at late times. 
Specifically, CHJ2 initially follows the exact trajectory, then deviates, and finally partially regains 
the stationary solution. This behaviour is more pronounced in the case 
with $k_y = 4$, where for $t > 2T$ CHJ2 manages to follow closely the exact solution. 
To validate the reference solution, we also report the theoretical decay 
law $\exp(-k_y^2 \nu t)$, which agrees excellently with the NSHJ results.

It is important to note that reproducing the stationary state of the NS equations accurately, requires 
increased spatial resolution. This is due to the fact that the velocity fields are obtained as derivative quantities according to Eq.~\eqref{eq:definition_v}, while the NSHJ system evolves its variables using second-order derivatives. 
Higher resolution is therefore necessary to correctly capture the subtle variations in the stationary regime.

Taken together, these findings suggest a hybrid strategy in which a second-order 
approximation is employed to capture long-time trends, while higher-order truncations 
are used to resolve short-time dynamics. 
In principle, this approach could enhance the overall accuracy of CHJ, even though 
its computational cost would depend strongly on the Reynolds number.  
This sensitivity to the Reynolds number is common to all Carleman 
formulations of fluid dynamics, except CLB, at least up to $Re\approx100$~\cite{psiq_26}.


\section{Conclusions}\label{sec:conclusions}

In this work, we have exploited the quantum formulation of classical fluid 
dynamics via the Schr\"odinger Navier-Stokes (SNS) equations first proposed by 
Dietrich and Vautherin back in 1985, to develop a corresponding quantum algorithm.
 
While the SNS approach provides a unified mathematical framework, suitable 
for the development of quantum algorithms, in practice, the strong nonlinearity 
associated to the dissipative term and the quantum potential present 
a major hurdle for the development of the corresponding quantum algorithm.   
As a way out, we have resorted to the inverse Madelung formulation, decomposing 
the flow field into rotational and irrotational components, which leads 
to a Navier--Stokes--Hamilton--Jacobi system of equations.

Applying Carleman embedding to these equations, we performed simulations 
up to fourth order, showing that Carleman linearization provides a viable alternative
to existing formulations.

We studied the behaviour of Carleman procedure at increasing truncation order, showing that it 
improves accuracy only up to a crossover time $t_{\mathrm{cross}}$, beyond which 
the second-order truncation provides the most reliable long-time behaviour. 
Such crossover time is found to increase with the truncation order $N_C$ and 
decrease with the Reynolds number.
At comparatively long times (a few dissipative time scales $T$) the 
second-order approximation correctly captures the decay trend, although 
its accuracy deteriorates at larger Reynolds numbers. 
A noticeable technical advance of this work is the introduction of a tensor-network 
strategy for Carleman simulations which permits to reduce the computational complexity 
by representing each term as a sum of rank-one tensors, taking the complexity from 
$\mathcal{O}\!\left((4G)^{N_C}\right)$ to \\$\mathcal{O}\!\left((N_t\,4G)^{N_C-2}(N_C-2)!\right)$, thereby enabling efficient simulations up to fourth order. 
This tensor-network technique is general and could be applied to Carleman linearization of other
fluid formulations, particularly Carleman Lattice Boltzmann, thereby suggesting a promising 
avenue for reducing resource requirements in quantum simulations of fluid systems.

Summarising, this paper shows that a fourth avenue to Carleman linearization of fluid dynamics
is available, inspired by quantum-like wave formulation of fluid 
mechanics and realised through a Hamilton-Jacobi representation.
To the best of our knowledge, this is the first quantum algorithm based on a wave
formulation of the genuine Navier-Stokes equations, including pressure, dissipation
and vorticity.

Deciding which one of the four (Navier-Stokes, Grad, Lattice Boltzmann, Hamilton-Jacobi)
is the most suitable in terms of convergence and accuracy is
a fairly complicated question, which depends on many aspects, not
least the details of the discrete formulation, as well as on the physical problem at hand.
The CLB shows an edge in terms of low-sensitivity to the Reynolds number, at least up to
$Re \sim 100$, quite likely due to its peculiar structure: exact streaming instead of finite
differencing of advection, and local collisions instead of Laplacian dissipation.
On the other hand, CLB gains these favorable properties at the expense of a larger
number of fields, typically $9$ in two dimensions (and generally 18 for thermal flows,  \cite{THLB} ) against $4$ for CHJ and $3$ for CNS,
which implies a surplus of memory at high orders of truncation as well as less sparsity.   
Extensive benchmarks are required to clarify these matters and advance the search of a quantum algorithm for fluid flows.


\section{Acknowledgments}
We acknowledge Simona Perotto for the valuable discussions. 
Furthermore, L.C. and S.S.  acknowledge financial support form the Italian 
National Centre for HPC, Big Data and Quantum Computing (CN00000013).
The authors have no conflicts to disclose. 
The data that support the findings of this study are available from the corresponding 
author upon reasonable request.

\bibliographystyle{apsrev4-2}
\bibliography{bibliography}
\appendix

\section{Explicit CHJ Matrices}\label{suppl:matrices}
 Let us define the discrete divergence operator $D=D_x+D_y$ and Laplace operator $L = D_{xx} + D_{yy}$.
In a 2D scenario, the expression for the matrix conveying the linear term in Eq.~\eqref{eq:structure j} is
\begin{equation}\label{eq:matrix_A_carleman}
    A_{\alpha \beta} = \mathbb{I} + \Delta t
    \begin{pmatrix}
        0 &0 &0 & 0
        \\-c_s^2\mathbb{I} & \nu L & \nu D_x & \nu D_y 
        \\ 0 &0 &\nu D_{yy} & -\nu D_{xy}
        \\ 0 &0 & -\nu D_{xy} & \nu D_{xx}
    \end{pmatrix} \, ,
\end{equation}

while the nonlinear terms are given by the tensor $B$
    \begin{equation}\label{eq:matrix_B1}
    \resizebox{.85\columnwidth}{!}{$
        B_{1} = \Delta t
        \begin{pmatrix}
            0 &-\Pi (D\otimes D + \mathbb{I}\otimes L) &0 & -\Pi (\mathbb{I}\otimes D_y + D_y\otimes\mathbb{I})
            \\ 0 &0 &0 &0
            \\ 0 &0 &0 &0
            \\ 0 &0 &0 &0
        \end{pmatrix}
        $
    }
    \end{equation}
    
    \begin{equation}\label{eq:matrix_B2}
    \resizebox{.85\columnwidth}{!}{$
        B_{2} = \Delta t
        \begin{pmatrix}
            0 &0 &0 &0
            \\ 0 &-\Pi(\frac{1}{2}D\otimes D) &\Pi(-D\otimes\mathbb{I}) &-\Pi(D_y\otimes\mathbb{I})
            \\ 0 &0 &-\Pi(\frac{1}{2}\mathbb{I}\otimes\mathbb{I}) &0
            \\ 0 &0 &0 &-\Pi(\frac{1}{2}\mathbb{I}\otimes\mathbb{I})
        \end{pmatrix}
        $}
    \end{equation}
    
    \begin{equation}\label{eq:matrix_B3}
        B_{3} = \Delta t
        \begin{pmatrix}
            0 &0 &0 &0
            \\ 0 &0 &-\Pi(D_y\otimes D_y) &\Pi(D_y\otimes D_x)
            \\ 0 &0 &0 &-\Pi(D_y \otimes \mathbb{I})
            \\ 0 &0 &0 &-\Pi(\mathbb{I} \otimes D_x)
        \end{pmatrix}
    \end{equation}
    
    \begin{equation}\label{eq:matrix_B4}
        B_{4} = \Delta t
        \begin{pmatrix}
            0 &0 &0 &0
            \\ 0 &0 &\Pi(D_x\otimes D_y) & -\Pi(Dx\otimes D_x)
            \\ 0 &0 &-\Pi(D_y \otimes \mathbb{I}) & -\Pi(\mathbb{I} \otimes D_x)
            \\ 0 &0 &0 &0
        \end{pmatrix} \,.
    \end{equation}

In the previous equations, $\Pi \in \mathbb{R}^{G \times G^2}$ is a rectangular operator that acts as a diagonal projector. It plays a key role in reducing the dimensionality from $J^{(2)} \in \mathbb{R}^{G^2}$ to match that of $J^{(1)} \in \mathbb{R}^G$, where $G$ again denotes the number of grid points.
The matrices $B_i$ contain the components that contribute to the evolution of the term $J_i$ in Eq.~\eqref{eq:J_carleman}, and are all part of the tensor B.
To illustrate this with an explicit example, consider the case $i=1$, where the first field corresponds to the density. 
Then, Eq.~\eqref{eq:structure j} reduces to
\begin{equation}
    \hat{\rho} = B_{1 \alpha \beta} \, J^{(2)}_{\alpha \beta} \,,
\end{equation}
where the corresponding row in the $A$ matrix is zero.

It should be noted that the matrices $B$ are formally defined on the non-local tensor space $J(\mathbf{r}_1)\otimes J(\mathbf{r}_2)$, which has dimension $16 G^2$, where $G$ is the number of grid points. In the original formulation, non-local couplings arise exclusively through finite-difference shifts. For instance, consider a one-dimensional derivative along $x$; terms of the form
$$
\frac{(\rho_{i+1}-\rho_{i-1})\chi_i}{2\Delta x}
$$
are formally non-local and would, in principle, require consideration of the full tensor $J^{(2)}$, rather than only the local contributions $J(\mathbf{r})\otimes J(\mathbf{r})$. Introducing the finite-difference operator
$$
D_x \rho_i = \frac{\rho_{i+1}-\rho_{i-1}}{2\Delta x},
$$
these contributions can be expressed as $(D_x \rho_i)\chi_i$. As a result, although the full non-local tensor structure of $B J^{(2)}$ must be retained, the physically relevant contributions entering the final expressions correspond to the diagonal components of $B J^{(2)}$. This projection is necessary to correctly account for the non-local couplings induced by finite differences while maintaining consistency with the original operator formulation.

\subsection{Block encoding}
We now briefly explain how we perform the block encoding of the update matrix used in Sec.~\ref{sec:quant_alg} of the main text. First we explain how to perform the block encoding of the matrix performing the update of the $J^{(1)}$ component
\begin{equation}
M_1=A\otimes\mathbb{I}\otimes\mathbb{I}+\overline{B}\;,
\end{equation}
where $\overline{B}$ is a square matrix that embeds $B$ in the top right sub-matrix and is zero otherwise. It is helpful to first find a block encoding for the shifted matrix
\begin{equation}
A\otimes\mathbb{I}\otimes\mathbb{I}+\overline{B} - \mathbb{I}\otimes\mathbb{I}\otimes\mathbb{I}\;,
\end{equation}
and than use this to generate the full matrix. From the explicit definitions provided above we know that the matrix $(A-\mathbb{I})$, and thus is extension to the full space present in $M_1$, is a sparse matrix with sparsity at most 10 in both rows and columns. We can also bound the largest matrix element in absolute value with
\begin{equation}
\mu_{(A-\mathbb{I})} = \Delta t \max\left[c_s^2,\frac{4\nu}{\Delta^2_x},\frac{\nu}{2\Delta_x}\right]=O\left(\frac{\nu\Delta t}{\Delta_x^2}\right)\;,
\end{equation}
with the second equality holding in the limit of sufficiently small spatial resolution. 

In a similar way we see that $B$, and thus $\overline{B}$, is a sparse matrix with maximum row and column sparsity at most 25 and matrix elements bounded from above by
\begin{equation}
\mu_{\overline{B}}=\mu_B=\frac{\Delta t}{\Delta x}\max\left[\frac{4}{\Delta_x},\frac{1}{2}\right]=O\left(\frac{\Delta t}{\Delta x^2}\right)\;,
\end{equation}
in the same limit used for $\mu_{(A-\mathbb{I})}$ above.

\begin{figure*}
    \centering
    \includegraphics[width=\linewidth]{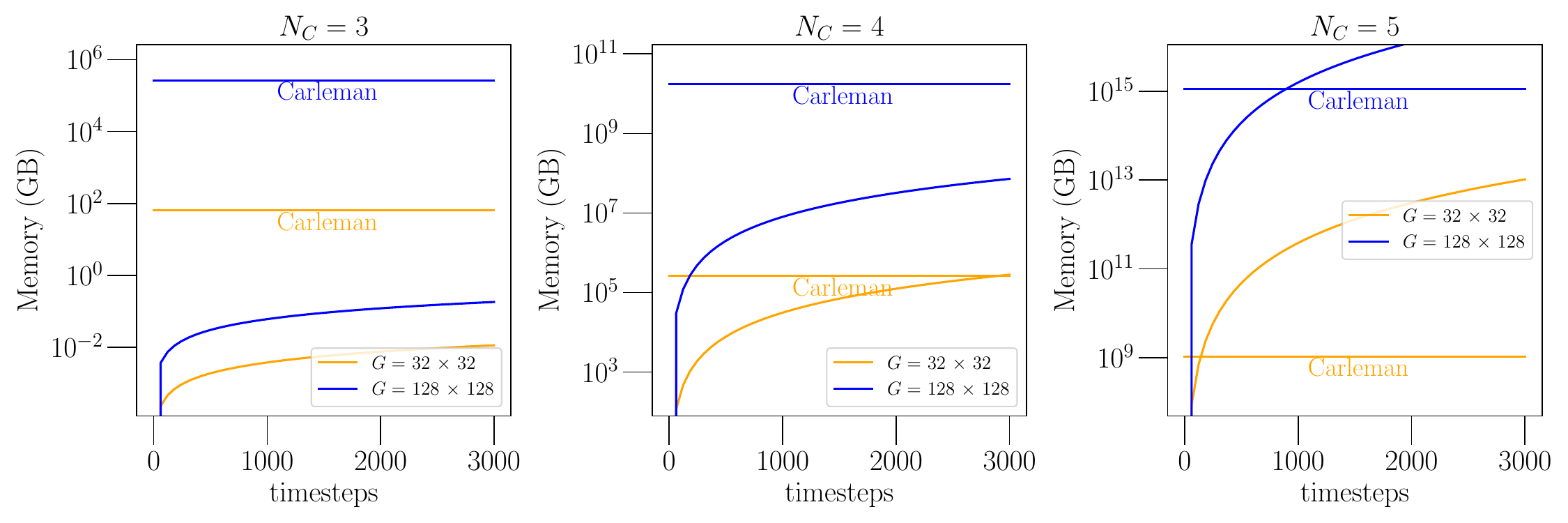}
    \caption{Comparison of scaling law in terms of memory required for the standard Carleman and tensor-network approaches. We report the behavior for different truncation orders $N_C$ and grid sizes $G$. The horizontal line represents the cost to implement the standard Carleman tensors, while the curved line indicates the memory requirement for the tensor-network approach. The advantage of the tensor network approach is evident for lower truncation orders and higher resolutions. }
    \label{fig:tn_scaling_price}
\end{figure*}

Using these estimates we can use find a block encoding with precision $\epsilon_1$ of the rescaled matrix
\begin{align}
    \Lambda_1 = &\frac{1}{35\max[\mu_{(A-\mathbb{I})},\mu_B]}
    \\&\nonumber
    \left(A\otimes\mathbb{I}\otimes\mathbb{I}+\overline{B} - \mathbb{I}\otimes\mathbb{I}\otimes\mathbb{I}\right)\;,
\end{align}
where $35$ is a bound on the maximum sparsity of the full matrix, using standard sparse-matrix techniques (see e.g.~\cite{gilyen2019quantum}) requiring $3$ additional ancilla qubits to embed the matrix into a unitary and a number of resources, memory and gates, polynomial in $n$ and polylog in the error $\epsilon$.

We can then use this, together with a single additional ancilla qubit to block encode the rescaled matrix
\begin{equation}
\frac{M_1}{1+35\max[\mu_{(A-\mathbb{I})},\mu_B]}=\frac{\mathbb{I}+35\max[\mu_{(A-\mathbb{I})},\mu_B]\Lambda_1}{1+35\max[\mu_{(A-\mathbb{I})},\mu_B]}\;.
\end{equation}
We then see that the sub-normalization factor, defined in Eq.~\eqref{eq:block_encoding} of the main text, for the encoding of the $M_1$ matrix is
\begin{equation}
\alpha_{M_1}=1+35\max[\mu_{(A-\mathbb{I})},\mu_B]=1+O\left(\frac{\Delta t}{\Delta x^2}\right)\;.
\end{equation}

Using a similar strategy for the matrix updating the $J^{(2)}$ components
\begin{equation}
M_2 = \mathbb{I}\otimes A\otimes A\;,
\end{equation}
we first write the following decomposition
\begin{align}
M_2 =&
\mathbb{I}\otimes (A-\mathbb{I})\otimes (A-\mathbb{I})+\mathbb{I}\otimes (A-\mathbb{I})\otimes\mathbb{I}
\\\nonumber
&+\mathbb{I}\otimes \mathbb{I}\otimes (A-\mathbb{I})+\mathbb{I}\otimes\mathbb{I}\otimes\mathbb{I}\;.
\end{align}
The second and third terms can now be block encoded exactly as we did before for $(A-\mathbb{I})$ matrix on the first register. The first term instead can be encoded as a sparse matrix from scratch using the fact that its sparsity is bounded by $100$ and its matrix elements are bounded in absolute value by $\mu^2_{(A-\mathbb{I})}$. The full block encoding can be done now using $5$ qubits, instead of the $4$ used before, for the embedding and a number of resources scaling similarly to the ones needed for the $M_1$ matrix. With the block encoding just described the sub-normalization factor takes the form
\begin{align}
\alpha_{M_2} 
&= 1+20\mu{(A-\mathbb{I})}+100\mu^2_{(A-\mathbb{I})}
\\&\nonumber=\left(1+10\mu{(A-\mathbb{I})}\right)^2\;.
\end{align}

In summary, the full block encoding of the evolution operator $M_2M_1$ can be done with $9$ ancilla qubit for the embedding and with a sub-normalization factor given by the product $\alpha_{M_1}\alpha_{M_2}=1+O(\Delta t/\Delta x^2)$. 

The success probability for applying this to a state $\ket{\Psi(t)}$ at time $t$ can then be written as
\begin{align}
\label{eq:succ_prob_app}
p_s &= \frac{\bra{\Psi(t)}M_1^\dagger M_2^\dagger M_2M_1\ket{\Psi(t)}}{\alpha^2_{M_1}\alpha^2_{M_2}}
\\&\nonumber
=\left\|\ket{\Psi(t+\Delta t)}\right\|^2\left(1+O\left(\frac{\Delta t}{\Delta x^2}\right)\right)\;.
\end{align}

\section{Tensor Network Representation}\label{suppl:TN_representation}
Consider the explicit $3-$rd order Carleman system
\begin{equation}
\begin{aligned}
\hat{J}^{(1)} &= A J^{(1)} + B J^{(2)}, \\
\hat{J}^{(2)} &= A^2 J^{(2)} + (AB + BA) J^{(3)}, \\
\hat{J}^{(3)} &= A^3 J^{(3)} 
\end{aligned}
\end{equation}
Here, $J^{(3)}$ can be effectively represented by a rank--1 tensor, since its evolution is governed solely by the term $A^{\otimes 3}$, which preserves factorization throughout the dynamics. This is no longer true for $J^{(2)}$, as its evolution includes contributions from $J^{(3)}$, making it non--separable in general. Nevertheless, rather than explicitly summing these contributions, $J^{(2)}$ can be interpreted as a superposition of factorized tensors.

To illustrate this mechanism, let us consider the first time step. Since $J^{(3)}$ is fully factorized, we write it as $t \otimes t \otimes t$, where $t$ contains all the information required to represent $J^{(3)}$. In the update of $J^{(2)}$, the contribution $(A \otimes B) J^{(3)}$ can then be written as $(A t) \otimes (B(t \otimes t))$. At the same time, the term $(A \otimes A) J^{(2)}$ preserves the factorized structure of $J^{(2)}$. As a result, after a single iteration, $\hat{J}^{(2)}$ can be expressed as a finite sum of rank-1 tensors, schematically of the form
$$
(At, At), \quad (At, B(t \otimes t)), \quad (B(t \otimes t), At),
$$
which altogether correspond to six rank--1 contributions. At the subsequent time step, the operator $A \otimes A$ acts on each of these terms independently, again preserving their separability. This structure is maintained as long as the summation of the individual contributions is postponed.

The summation becomes unavoidable at the level of $J^{(1)}$, which is the quantity of direct physical interest. In this case, the operator $B$ acts on each factorized component of $J^{(2)}$, after which the resulting terms are summed. However, each contribution of $B$ can be written in the form $\mathrm{diag}(b_i \otimes b_j)$, allowing its action to be decomposed into separate multiplications by $b_i$ and $b_j$ on the first and second factors of $J^{(2)}$, respectively. Exploiting the symmetry of the system, this procedure allows the intermediate representations to remain factorized.

This approach avoids the explicit construction of full tensors. 
As a consequence, the computational complexity no longer grows exponentially with the truncation order, but only linearly with the number of retained factorized components. For a third-order truncation, where the NSHJ dynamics requires the full tensor $J(x) \otimes J(y)$, the naive cost scales as $(4G)^2$, while the tensor network formulation leads to a cost of $(1 + 2N_t)\, 2(4G)$, where $G$ is the number of grid points and $N_t$ the number of time steps. In both cases, $J^{(3)}$ is kept implicitly factorized.

The advantage becomes even more pronounced for a fourth-order truncation. In this case, the cost scales as 
\begin{equation}
(1 + 2 N_t N_{J^{(3)}} + N_t N_{J^{(4)}})\, 2(4G)\, ,    
\end{equation}
with $N_{J^{(3)}} = (1 + 3N_t)\, 3(4G)$ and $N_{J^{(4)}} = 1$, leading to an overall complexity of
\begin{equation}
    (1 + 2N_t((1 + 3N_t)(3(4G))) + N_t)\, 2(4G) \, .    
\end{equation}
From this expression, one can extract that, for a truncation of order $N_C$, the dominant contribution scales as
$$
\mathcal{O}\!\left((N_t\, 4G)^{N_C-2} (N_C-2)!\right),
$$
to be compared with the $\mathcal{O}((4G)^{N_C})$ scaling of the full tensor formulation. These scaling relations are presented in Fig.~\ref{fig:tn_scaling_price} for $N_C = \{3,4,5\}$ and for different grid size $G$. It is evident how the tensor network approaches show its advantages with both low $N_C$ and increasing grid size. As the truncation order grows, in order to have advantages with the tensor network approach the grid size needs to grow, this is explicit in the third column of Fig.~\ref{fig:tn_scaling_price}.


\end{document}